\documentclass[12pt,pre,aps,preprint,showpacs]{revtex4-1}

\usepackage{graphicx}
\usepackage{amsfonts}
\usepackage{amsmath}
\usepackage{mathtools}
\usepackage{float}
\usepackage{epsfig}
\usepackage{color}
\usepackage{etoolbox}
\usepackage{gensymb}
\usepackage{multirow}
\usepackage[mathscr]{euscript}
\usepackage{pifont}

\newcommand{\ee}[1]{\ensuremath{10^{#1}}}

\begin{document}

\title{Precise algorithm to generate random sequential adsorption of hard polygons at saturation}

\author{G. Zhang}

\email{gezhang@alumni.princeton.edu}

\affiliation{\emph{Department of Physics}, \emph{University of Pennsylvania}, Philadelphia, Pennsylvania, 19104, USA }

\begin{abstract}
Random sequential adsorption (RSA) is a time-dependent packing process, in which particles of certain shapes are randomly and sequentially placed into an empty space without overlap.  In the infinite-time limit, the density approaches a ``saturation'' limit. Although this limit has attracted particular research interest, the majority of past studies could only probe this limit by extrapolation. 
We have previously found an algorithm to reach this limit using finite computational time for spherical particles, and could thus determine the saturation density of spheres with high accuracy.
In this paper, we generalize this algorithm to generate saturated RSA packings of two-dimensional polygons.
We also calculate the saturation density for regular polygons of three to ten sides, and obtain results that are consistent with previous, extrapolation-based studies.
\end{abstract}

\maketitle

\section{Introduction}

Random sequential adsorption (RSA) \cite{feder1980random}, also called random sequential addition \cite{widom_1966_rsa}, is a stochastic process widely used to model a variety of physical, chemical, and biological phenomena, including structure of cement paste \cite{xu2013numerical}, ion implantation in semiconductors \cite{roman1983computer}, protein adsorption \cite{feder1980adsorption}, particles in cell membranes \cite{finegold1979maximum}, and settlement of animal territories \cite{tanemura1980geometrical}. Starting from a large, empty region in $d$-dimensional Euclidean space, particles of certain shapes are randomly and sequentially placed into the volume subject to a nonoverlap constraint: New particles are kept only if they do not overlap with any existing particles, and are discarded otherwise. 
One can stop this process at any time, obtaining configurations with time-dependent densities. As time increases, the density approaches a ``saturation'' or ``jamming'' limit, $\phi_s$. 

The RSA process of various particle shapes have been studied, including spheres in one through eight dimensions \cite{renyi1963one, cooper_1987_RSA, wang1994fast, wang1998random, torquato2006random, zhang2013precise}, squares and rectangles \cite{vigil1989random, vigil1990kinetics, viot1990random, brosilow1991random,  viot1992random}, 
polygons \cite{ciesla2014random},
ellipses \cite{talbot1989unexpected, viot1992random, sherwood_1999_RSA_elipse},
disk polymers \cite{ciesla2015shapes}, 
cubes \cite{ciesla2018random},
spheroids \cite{sherwood_1999_RSA_elipsoid}, superdisks \cite{gromenko2009random}, sphere polymers \cite{ciesla2013modelling, ciesla2013random2},
and four-dimensional hypercubes \cite{blaisdell1982random}.
For non-spherical shapes, particle orientations may be random or fixed.
Although previous researchers have studied a myriad of combinations of space dimensions, shapes, and orientations, the determination of $\phi_s$ has always been of particular interest. However, doing so is also particularly difficult since one cannot afford infinite computational time to reach the saturation limit. To overcome this problem, a very common strategy is to find out finite-time densities and then to extrapolate to the infinite-time limit \cite{cooper_1987_RSA, vigil1989random, talbot1989unexpected, vigil1990kinetics, viot1992random, sherwood_1999_RSA_elipse, sherwood_1999_RSA_elipsoid, torquato2006random, ciesla2013modelling, ciesla2013random2, ciesla2014random, ciesla2015shapes, ciesla2018random}. 

Instead of extrapolation, $\phi_s$ for some systems can be ascertained by other strategies. For one-dimensional rods, analytical calculations found $\phi_s=0.7475979202\ldots$ \cite{renyi1963one}. 
For disks, spheres, and $d$-dimensional hyperspheres, we have previously found a numerical algorithm to reach the saturation limit with finite computational time \cite{zhang2013precise}. 
The algorithm takes advantage of the fact that when generating RSA packings of spheres of radius $R$, the distance between any two sphere centers cannot be smaller than $2R$. The part of space that is not within $2R$ distance to any existing sphere center is called ``available space,'' since a new sphere will be kept if and only if its center falls inside the available space. Thus, one can avoid insertion attempts in the unavailable part of the space \cite{wang1994fast, wang1998random, brosilow1991random, torquato2006random, zhang2013precise}.
Saturation can be achieved by gradually increasing the resolution as additional spheres are inserted, eventually eliminating all available spaces \cite{wang1994fast, zhang2013precise}. Specifically, our algorithm consists of the following steps:
\begin{enumerate}
\item Perform a certain number of trial insertions to generate a near-saturation configuration. \item Divide the simulation box into voxels ({\it i.e.,} $d$-dimensional pixels) with side lengths comparable to $R$. Some voxels are completely covered by an existing sphere of radius $2R$, and therefore cannot contain any available space. They are excluded from the voxel list.
\item Perform a certain number of trial insertions inside the remaining voxels.
\item Divide each voxel into $2^d$ subvoxels by cutting it in half in each direction, and find possibly available subvoxels. 
\item The process of trial insertions and voxel division is repeated until the number of available voxels reaches zero, at which point we know that saturation is guaranteed since we only discard voxels that cannot contain any available space.
\end{enumerate}

Reference~\onlinecite{zhang2013precise} ended with a proposal to extend this algorithm to generate saturated RSA packings of nonspherical shapes with random orientations. In this case, whether an incoming particle overlaps with existing ones depends on not only its location but also its orientation. For a $d$-dimensional particle with $d_f$ rotational degrees of freedom, one can construct a $(d+d_f)$-dimensional auxiliary space. Each point in this space would correspond to a trial insertion at a particular location with a particular orientation. One could thus use voxels to track the available parts of this higher dimensional auxiliary space and generate saturated RSA packings. However, Ref.~\onlinecite{zhang2013precise} did not propose any method to test for voxel availability, which is a nontrivial task.

In this paper, we use this idea to generate saturated RSA packings of 2D polygons with random orientations. We present a way to test for voxel availability based on worst-case error analysis in Sec.~\ref{sec:VoxelAvailability}. We find that occasionally, the generalization of this algorithm needs a special tweak, detailed in Sec.~\ref{sec:problem}. In Sec.~\ref{sec:result}, we use this algorithm to find $\phi_s$ for regular polygons, which generally increases as the number of sides increases and approaches $\phi_s$ for disks.

\section{Algorithmic Details}
\label{sec:algorithm}

In order to use the algorithm described in Ref.~\onlinecite{zhang2013precise}, one need to supplement two subroutines: one to determine if two particles are overlapping and another to prove that certain voxels that cannot contain any available space. Here we describe these two subroutines.

\subsection{Polygon overlap test}
To test if two polygons overlap, we first perform simple tests using their inscribed circles and circumscribed circles: If the inscribed circles overlap, then the two polygons must overlap. If the circumscribed circles do not overlap, then the two polygons cannot overlap. If these two simple tests fail to find a definitive answer, we then test if any two sides of the two polygons intersect with the following theorem \cite{lineIntersect}:

Define $O(\mathbf r_1, \mathbf r_2, \mathbf r_3)=(y_2-y_1)(x_3-x_2)-(x_2-x_1)(y_3-y_2)$, where $\mathbf r_i=(x_i, y_i)$ is a two-dimensional point, and then two line segments $(\mathbf r_1, \mathbf r_2)$ and $(\mathbf r_3, \mathbf r_4)$ intersect if and only if
\begin{align}
&O(\mathbf r_1, \mathbf r_2, \mathbf r_3)O(\mathbf r_1, \mathbf r_2, \mathbf r_4)<0\mbox{, and}\\
&O(\mathbf r_3, \mathbf r_4, \mathbf r_1)O(\mathbf r_3, \mathbf r_4, \mathbf r_2)<0.
\end{align}

\subsection{Voxel availability test}
\label{sec:VoxelAvailability}
Our test for voxel availability is based on the aforementioned polygon-overlap test.
Since a two-dimensional polygon has 1 rotational degree of freedom, the voxels are three-dimensional. Let $(x, y)$ be the location of the center of a polygon and let $\theta$ be the angle between the orientation of the polygon and some reference orientation. A point in the voxel space can be represented by $(x, y, \theta)$, while a voxel centered at this point can be represented as $(x\pm \delta x, y\pm \delta y, \theta\pm \delta \theta)$, where $\delta x$, $\delta y$, and $\delta \theta$ are a half of the side length of the voxel in each direction. With this formulation, a voxel can be interpreted as a collection of trial insertions near $(x, y, \theta)$ with some error bounds $\delta x$, $\delta y$, and $\delta \theta$. To prove that a voxel cannot contain any available space ({\it i.e.,} to prove that such trial insertions always fail), we just need to perform a rigorous worse-case error analysis to prove that no matter how $x$, $y$, and $\theta$ vary in the ranges $[x-\delta x, x+\delta x]$, $[y-\delta y, y+\delta y]$, and $[\theta-\delta \theta, \theta+\delta \theta]$, the upcoming particle will always overlap with an existing one. Specifically, let $(\mathbf r_1, \mathbf r_2)$ be a side of an existing polygon and let $(\mathbf r_3, \mathbf r_4)$ be a side of an upcoming polygon inside the voxel, then $\mathbf r_3$ and $\mathbf r_4$ carry uncertainties while $\mathbf r_1$ and $\mathbf r_2$ do not. Let $(l_3, \theta_3)$ and $(x_3, y_3)$ be the polar coordinates and Cartesian coordinates of $\mathbf r_3$, we have $x_3=x+l_3 \cos(\theta_3)$, and the associated error bound is
\begin{align}
\delta x_3 &= [x\pm \delta x + l_3 \cos(\theta_3\pm \delta \theta)] - [x+l_3 \cos(\theta_3)]\\
&\le \delta x+l_3\delta \theta.
\end{align}
Similarly,
\begin{align}
\delta y_3 \le \delta y+l_3\delta \theta.
\end{align}
For simplicity, we can require that a voxel always have equal side lengths in $x$ and $y$ directions, in this case $\delta x=\delta y$ and $\delta x_3=\delta y_3$. We define $\delta_3\equiv\delta x_3=\delta y_3$. The error bounds for the Cartesian coordinates of vertex 4 are similar:
\begin{align}
\delta x_4 \le \delta x+l_4\delta \theta\equiv\delta_4,
\end{align}
\begin{align}
\delta y_4 \le \delta y+l_4\delta \theta=\delta_4,
\end{align}
The associated worse-case error in the $O$ functions used in Eqs.~(1) and (2) are thus
\begin{align}
\delta O(\mathbf r_1, \mathbf r_2, \mathbf r_3) &= (y_2-y_1)(x_3\pm \delta_3-x_2)-(x_2-x_1)(y_3\pm \delta_3-y_2)\\&-(y_2-y_1)(x_3-x_2)-(x_2-x_1)(y_3-y_2)\\
&\le (|y_2-y_1| + |x_2-x_1|)\delta_3;
\end{align}
similarly
\begin{align}
\delta O(\mathbf r_1, \mathbf r_2, \mathbf r_4) \le (|y_2-y_1| + |x_2-x_1|)\delta_4
\end{align}
and
\begin{align}
\delta O(\mathbf r_3, \mathbf r_4, \mathbf r_1) &=(y_4\pm \delta_4 - y_3\pm\delta_3)(x_1-x_4\pm\delta_4)-(y_4-y_3)(x_1-x_4)\\&+(x_4\pm \delta_4 - x_3\pm\delta_3)(y_1-y_4\pm\delta_4)-(x_4-x_3)(y_1-y_4)\\
&\le (\delta_3+\delta_4)(|x_1-x_4|+|y_1-y_4|+2\delta_4)+\delta_4(|y_4-y_3|+|x_4-x_3|);
\end{align}
similarly
\begin{align}
\delta O(\mathbf r_3, \mathbf r_4, \mathbf r_2) \le (\delta_3+\delta_4)(|x_2-x_4|+|y_2-y_4|+2\delta_4)+\delta_4(|y_4-y_3|+|x_4-x_3|).
\end{align}
These error bounds allow us to prove certain voxels' unavailability: If Eqs.~(1) and (2) hold, and if each $O$ function's error bound is smaller than its absolute value, then we know these $O$ functions cannot change sign no matter how $x$, $y$, and $\theta$ vary within their respective limit, and Eqs.~(1) and (2) will always hold. We thus proved that the voxel cannot contain available space.

It is noteworthy that the errors could be smaller than the worse-case bounds we derived. Thus, a completely unavailable voxel could be miscategorized as an available one. This is nevertheless not a problem for two reasons: First, such miscategorization can cause us to retain unavailable voxels but can never cause us to discard available ones. 
Second, as we repeatedly divide the voxels and drive all error bounds to zero, such miscategorization will eventually disappear. 
The ending configuration is thus still guaranteed to be saturated. 

\subsection{An unexpected problem and its solution}
\label{sec:problem}
With the aforementioned subroutines supplementing the split-voxel algorithm, we are ready to generate saturated RSA configurations of 2D polygons.
However, in doing so for 2D squares and regular hexagons, we found an unexpected problem: The number of voxels occasionally grows to extremely large numbers ($>\ee{8}$) for moderately-sized systems ($\approx 3000$ particles). Sometimes the number of voxels suddenly drops to zero after becoming extremely large, but sometimes our program crashes because of insufficient memory before the drop could happen. This is in contrast with the situation for disks, equilateral triangles, and regular pentagons, where the number of voxels always decays smoothly as they are split.

\begin{figure}
\begin{center}
\includegraphics[width=0.8\textwidth]{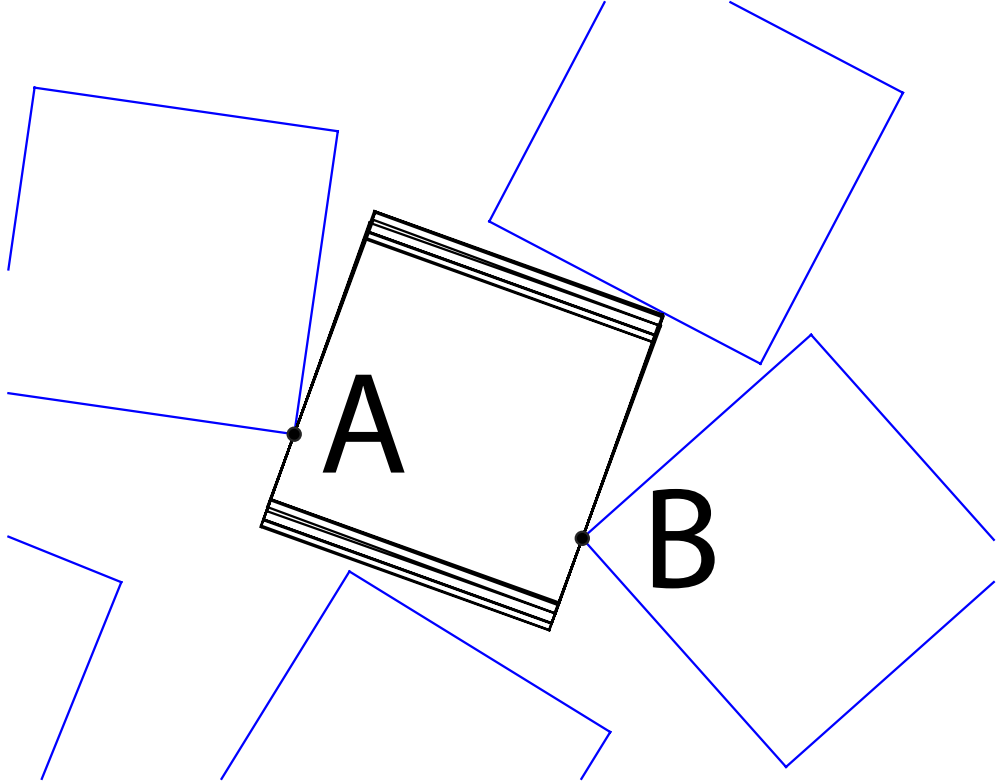}
\end{center}
\caption{Plot of 10 randomly selected voxel centers when the voxel-number-explosion problem happened when generating a saturated RSA packing of squares. Note that a voxel center has a three-dimensional coordinate $(x, y, \theta)$ and represents a trial insertion at location $(x, y)$ and orientation $\theta$. Hence, we use a black square to represent a voxel center. Blue squares are adjacent existing particles. The distance between points A and B is 0.999 992 times the side length of a square. Therefore, inserting a new square at the place indicated by these voxels is impossible.}
\label{Problem}
\end{figure}

To understand the reason, we plotted 10 randomly selected voxel centers when the problem happened in Fig.~\ref{Problem}. Surprisingly, all of the selected voxels are concentrated in a very small part of the configuration. In Fig.~\ref{Problem}, the distance between points A and B is 0.999 992 times the side length of a square. Therefore, inserting a new square at this place is impossible. Nevertheless, the algorithm could not realize this impossibility until voxel-space resolution becomes extremely fine. Figure \ref{Problem} also indicates that this problem can only occur when the polygon has at least one pair of parallel sides, and therefore explains why we observed this problem only for certain shapes.


Our solution to this problem is to run very deep tests of voxel availabilities. Specifically, we define a level-0 test of voxel availability as our original test outlined in Sec.~\ref{sec:VoxelAvailability}.
We define a level-$n$ ($n>0$) test of voxel availability as follows:
\begin{enumerate}
\item If the voxel can be proved unavailable with the procedure outlined in Sec.~\ref{sec:VoxelAvailability}, then declare the voxel unavailable.
\item Otherwise, if the voxel center represents an incoming particle that does not overlap with existing particles, declare the voxel available.
\item Otherwise, divide the voxel into $2^3$ subvoxels and run a level-$(n-1)$ test of each subvoxel. If any subvoxel is available, declare the original voxel available.
\item Otherwise, declare the original voxel unavailable.
\end{enumerate}
The deep test retains a desired property of the original voxel availability test: An unavailable voxel may be misjudged as an available one if $n$ is finite, but an available voxel will never be misjudged as an unavailable one. 
Therefore, one can safely employ a deep test on voxels and remove unavailable ones, without worrying about discarding any available space. 
If the particle has a pair of parallel sides, we randomly sample 100 voxels each time a voxel list is generated. If at least 50 of them are within a distance of $R_{ins}$, the radius of the inscribed circle of a particle, then this problem is suspected. We run a level-4 check on all voxels and discard unavailable ones. If all of the sampled voxels are within a distance of $R_{ins}$, then this problem is strongly suspected, and we run a level-12 check on all voxels and discard unavailable ones.

\begin{figure}
\begin{center}
\includegraphics[width=0.4\textwidth]{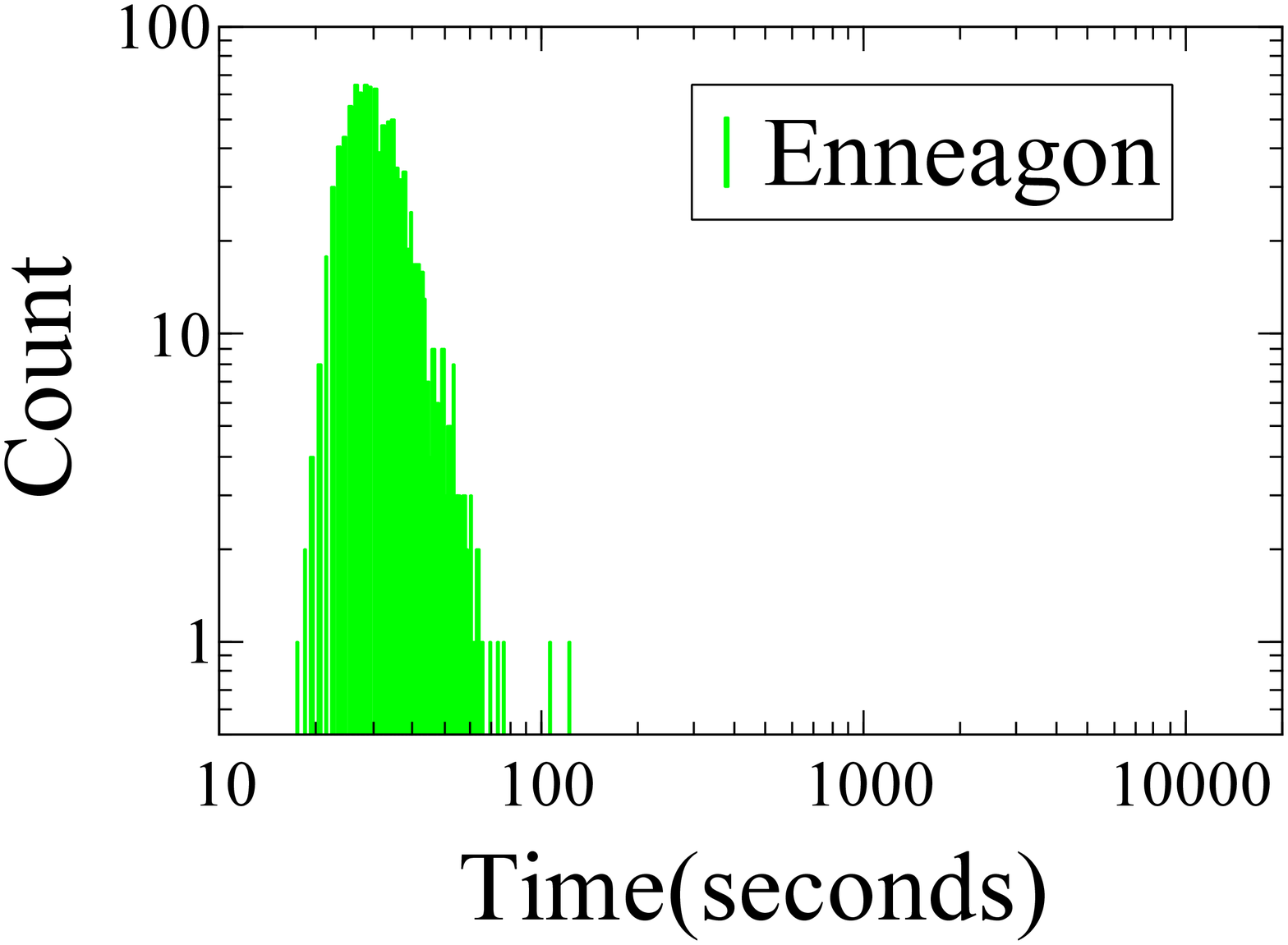}
\includegraphics[width=0.4\textwidth]{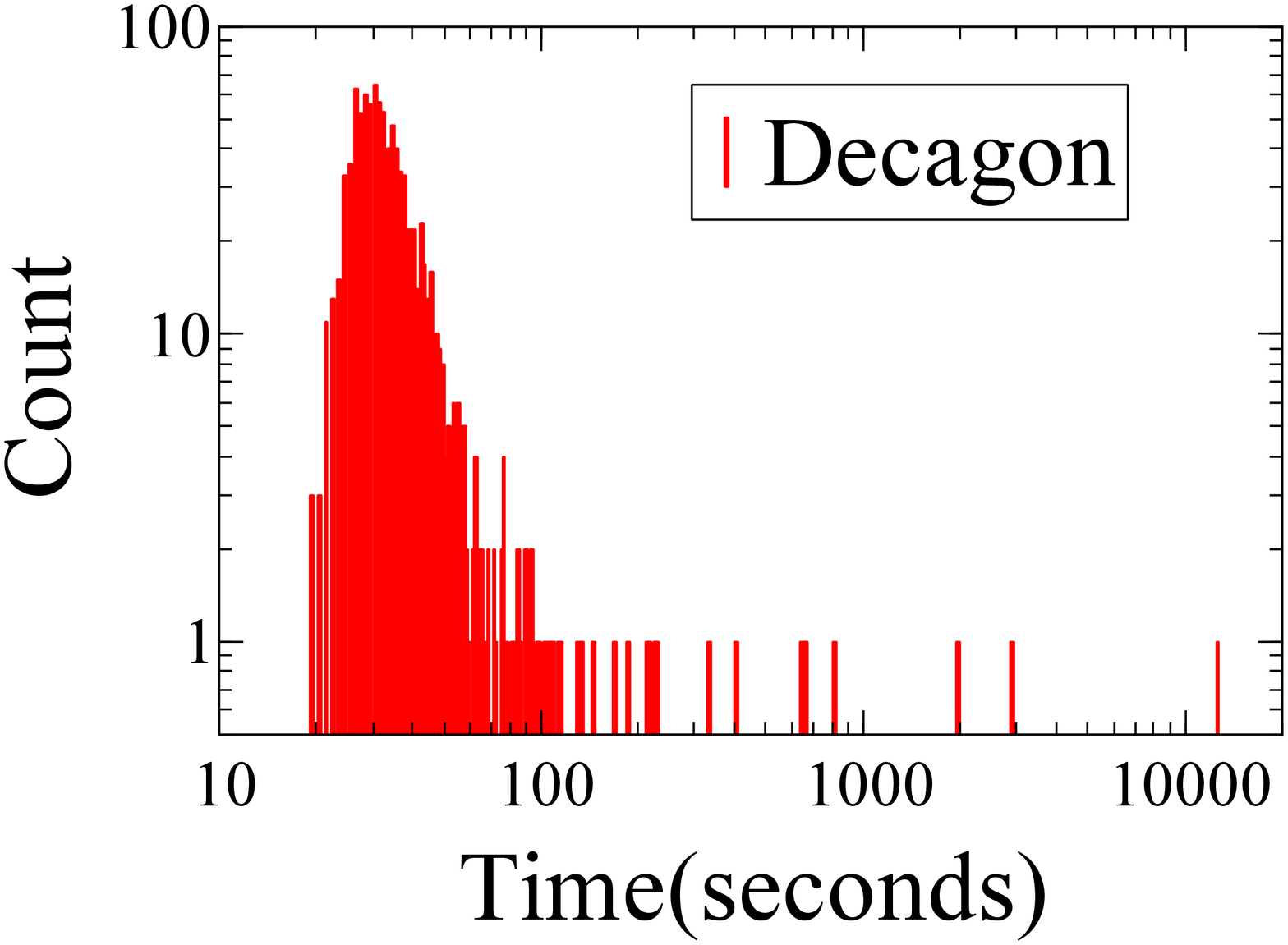}
\end{center}
\caption{Histogram of the time taken to generate an RSA configuration of (left) enneagons and (right) decagons of size $L_{\mbox{particle}}/L_{\mbox{box}}=0.01$ on a computer with an Intel Xeon E5-2696 v3 central processing unit.}
\label{histogram}
\end{figure}

The deep test successfully solves this problem but can be very time consuming. To illustrate this point, we show the histogram of the time taken to generate a saturated RSA configuration of enneagons and decagons in Fig.~\ref{histogram}. Only decagons are susceptible to this problem. The time distribution for decagons resembles that for enneagons, except that the former also exhibit a very long tail on the long-time side, corresponding to the extra time needed for the deep test.

One could argue that as Fig.~\ref{Problem} shows, this problem only happens in saturated locations. Thus, the simplest ``solution'' would be to just declare the configuration saturated whenever the problem is detected. If a strict proof of saturation is not required, this simple ``solution'' may be desirable, especially since deep testing voxel availability is very time-consuming. However, we choose the deep-test solution since in this work, we want to ensure that each configuration is saturated.

\section{Results and Discussion}
\label{sec:result}
To demonstrate the correctness and usefulness of this algorithm, we generate saturated RSA configurations of regular polygons, and compare $\phi_s$ with previous results. For each particle shape, we generate 1000 configurations with system size $L_{\mbox{particle}}/L_{\mbox{box}}=0.01$, 100 configurations with $L_{\mbox{particle}}/L_{\mbox{box}}=0.003$, and 10 configurations with $L_{\mbox{particle}}/L_{\mbox{box}}=0.001$. Here $L_{\mbox{particle}}$ is the distance between a particle's center and its vertex, and $L_{\mbox{box}}$ is the side length of the simulation box. The resulting saturation density is summarized in Table~\ref{tab:phis}.

\begin{table}[]
\centering
\caption{Saturation density $\phi_s$ for various particle shapes and system sizes. Here all error estimates are calculated from $\sigma = \sqrt{(<\phi_s^2>-<\phi_s>^2)/(N_c-1)}$, where $<\cdots>$ indicates averaging over all configurations, and $N_c$ is the number of configurations. }
\label{tab:phis}
\begin{tabular}{|c|c|c|c|}
\hline
\multirow{2}{*}{Shape} & \multicolumn{3}{l|}{$\phi_s$ for $L_{\mbox{particle}}/L_{\mbox{box}}=$} \\ \cline{2-4} 
                       & 0.01                   & 0.003                  & 0.001                 \\ \hline
Equilateral triangles  & $0.525892\pm0.000064$  & $0.525993\pm0.000058$  & $0.525820\pm0.000066$ \\ \hline
Squares                & $0.527719\pm0.000085$  & $0.527482\pm0.000080$  & $0.527594\pm0.000070$ \\ \hline
Regular pentagons      & $0.541319\pm0.000087$  & $0.541241\pm0.000088$  & $0.541344\pm0.000072$ \\ \hline
Regular hexagons       & $0.539114\pm0.000093$  & $0.539216\pm0.000087$  & $0.539060\pm0.000095$ \\ \hline
Regular heptagons      & $0.542143\pm0.000093$  & $0.542197\pm0.000093$  & $0.541959\pm0.000124$ \\ \hline
Regular octagons       & $0.542329\pm0.000094$  & $0.542494\pm0.000090$  & $0.542328\pm0.000098$ \\ \hline
Regular enneagons      & $0.544055\pm0.000092$  & $0.544044\pm0.000098$  & $0.544059\pm0.000089$ \\ \hline
Regular decagons       & $0.544104\pm0.000094$  & $0.544278\pm0.000100$  & $0.544259\pm0.000124$ \\ \hline
\end{tabular}
\end{table}

For all shapes, the difference in $\phi_s$ between different system sizes is comparable to the error estimate, suggesting negligible finite-size effect. Indeed, this is consistent with Ref.~\onlinecite{ciesla2017boundary}, which found minimal finite-size effect for even smaller ($L_{\mbox{particle}}/L_{\mbox{box}} \approx 0.2$) saturated RSA configurations with periodic boundary configurations. This is also expected in light of Ref.~\onlinecite{bonnier1994pair}, which found that the pair correlation functions of RSA configurations decay super-exponentially.
With negligible finite-size effect, we think the best estimate of $\phi_s$ for each shape can be obtained by simply averaging the results for different system sizes. This yields $\phi_s=0.525902\pm0.000036$ for triangles, $\phi_s=0.527598\pm0.000045$ for squares, $\phi_s=0.541301\pm0.000047$ for pentagons, $\phi_s=0.539130\pm0.000053$ for hexagons, $\phi_s=0.542100\pm0.000060$ for heptagons, $\phi_s=0.542383\pm0.000054$ for octagons, $\phi_s=0.544053\pm0.000054$ for enneagons, and $\phi_s=0.544214\pm0.000062$ for decagons.

Previous researches by extrapolating finite-time RSA densities found the saturation density of squares to be $0.523-0.532$ \cite{vigil1989random} and $0.530\pm0.001$ (reported in both Refs. \cite{viot1990random} and \cite{viot1992random}). Our result is within the former range but slightly below the latter (by two and a half times their error bar). Could this indicate a mistake, for example, that leads to the generation of unsaturated configurations? One way to double check is to calculate RSA saturation densities of regular $n$gons with large $n$, since as $n$ increases, $\phi_s$ should approach that for disks, $0.547067\cdots $ \cite{ciesla2017boundary}. We thus calculated $\phi_s$ for 19gons and 29gons, and found $0.546210\pm0.000080$ and $0.546701\pm0.000067$, respectively. These densities indeed approach $\phi_s$ for disks, and thus this does not suggest the existence of such a mistake.

\begin{figure}
\begin{center}
\includegraphics[width=0.8\textwidth]{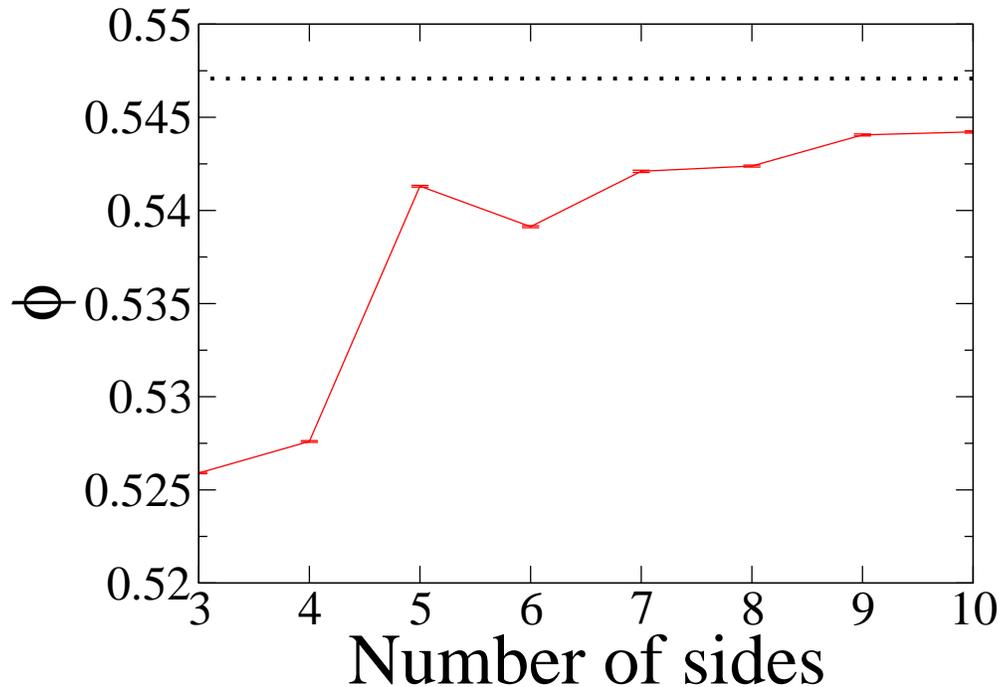}
\end{center}
\caption{(Red solid line) RSA saturation density for regular polygons, as a function of the number of sides. Note that each data point is associated with an error bar that is barely visible. (Black dotted line) RSA saturation density for disks. }
\label{phi}
\end{figure}

We plot $\phi_s$ versus the number of sides of the polygon in Fig.~\ref{phi}, which shows that $\phi_s$ increases as the number of sides increases except that $\phi_s$ for hexagons is lower than that for pentagons.
More generally, $\phi_s$ tend to be slightly higher than the trend when the number of sides is odd, and slightly lower otherwise. Overall, our results appear to be consistent with Ref.~\onlinecite{ciesla2014random}, which plotted (but not listed) $\phi_s$ for regular polygons obtained by infinite-time extrapolation.

\section{Conclusions}
To summarize, we have developed in this paper a generalization of the split-voxel algorithm described in Ref.~\onlinecite{zhang2013precise}, based on worst-case error analysis method. We support the correctness of this method by finding the RSA saturation densities of 2D regular polygons with three to ten sides, and verifying their consistency with previous results.

A program implementing this algorithm is available as Supplementary Material.

\begin{acknowledgments}
We thank the U.S. Department of Energy, Office of Basic Energy Sciences, Division of Materials Sciences and Engineering under Award DE- FG02-05ER46199. We also thank the UPenn MRSEC for computational support provided by the LRSM HPC cluster at the University of Pennsylvania.
\end{acknowledgments}

\bibliography{thesis}

\end{document}